\documentclass[dvips,11pt,a4paper]{article}
\usepackage[dvips,colorlinks,bookmarksopen,bookmarksnumbered,citecolor=red,urlcolor=blue]{hyperref}

\usepackage[T1]{fontenc}         
\usepackage[latin1]{inputenc}
\usepackage{graphicx}
\usepackage{amsmath}
\usepackage{bm}
\usepackage{tikz}
\usepackage{color}
\usepackage{helvet}
\usepackage{courier}
\usepackage{makeidx}
\usepackage{footmisc}
\usepackage{type1cm}         
\usepackage{textcomp}
\usepackage{subfig}
\usepackage{enumitem,xcolor}    

\DeclareCaptionLabelSeparator{colon}{. }
\usepackage[labelfont={bf},labelsep={colon}]{caption}

\makeindex

\DeclareMathAlphabet\mathbfcal{OMS}{cmsy}{b}{n}

\begin{document}
\thispagestyle{empty}
\bibliographystyle{plain}



\begin{center}
\textcolor{blue}{ \Large  \bf Dynamics of a Relativistic Particle in Discrete Mechanics } \\
\vspace{3.mm}
{\bf Jean-Paul Caltagirone } \\
\vspace{3.mm}
{ \small Universit{\'e} de Bordeaux  \\
   Institut de M{\'e}canique et d'Ing{\'e}ni{\'e}rie \\
   Département TREFLE, UMR CNRS n° 5295\\
  16 Avenue Pey-Berland, 33607 Pessac Cedex  \\
\textcolor{blue}{\texttt{ calta@ipb.fr }  } }
\end{center}

\small
{\bf Abstract}

The study of the evolution of the dynamics of a massive or massless particle shows that in special relativity theory, the energy is not conserved. From the law of evolution of the velocity over time of a particle subjected to a constant acceleration, it is possible to calculate the total energy acquired by this particle during its movement when its velocity tends towards the celerity of light. The energy transferred to the particle in relativistic mechanics overestimates the theoretical value.

Discrete mechanics applied to this same problem makes it possible to show that the movement reflects that of Newtonian mechanics at low velocity, to obtain a velocity which tends well towards the celerity of the medium when the time increases, but also to conserve the energy at its theoretical value. This consistent behavior is due to the proposed physical analysis based on the compressible nature of light propagation.

\normalsize

\vspace{3.mm}

\textcolor{blue}{\section{Introduction} }

The limitation of the velocity of a material medium or of a particle, with or without mass, to the celerity of light was introduced by H.A. Lorentz at the end of the 19th century. H. Poincar{\'e} gave an interpretation in 1904 based on the increase in inertia with velocity, but this limitation is attributed today to some purely cinematic reasons. This limitation was checked experimentally by W. Bertozzi \cite{Ber64} in 1964. For two observers, the $\gamma$ factor defined by Lorentz translates the dilation of time $dt = \gamma \: dt'$ with the $\gamma \ge 1$ property. Lorentz was led to introduce this factor in order to make Maxwell's equations invariant through a transformation, later called the Lorentz transformation by H. Poincar{\'e} and A. Einstein \cite{Lor04}. Many textbooks on the theory of special relativity describe the genesis of the remarkable advances made by these authors and many others, such as M. Planck.
The point of view adopted in relativistic mechanics was related to the presence of two observers in a three-dimensional space and the way in which each of them could describe the phenomena in the other frame of reference; this led A. Einstein to issue his principle on the invariance of the laws of physics for all inertial frames of reference. The time dilation and the length contraction are only interpreted in the concept related to the presence of two  frames of reference.

The point of view adopted in discrete mechanics is radically different: the observer is alone on an oriented edge  $\Gamma$ of length $d$. He can only record accelerations and cannot observe uniform motions, translation and rotation; the interactions with the vicinity defined by the celerity of the wave $c$ are of cause and effect \cite{Cal19a}. Lengths and velocities are second-order quantities that are updated from the acceleration $\bm \gamma$, the only absolute quantity. It is no longer necessary to consider the time dilation, the length contraction or the increase towards infinity of the apparent mass; the time runs regularly over discrete steps of amplitude $dt$. The approaches of relativistic mechanics and discrete mechanics will be confronted on the basis of an emblematic problem, that of the dynamics of a relativistic particle.

The goal is to calculate the proper acceleration of a particle subjected to a constant external acceleration $\mathbf g$ in a one-space dimension. This case very schematically simulates the behavior of charged particles in a linear accelerator. The theory of special relativity predicts that velocity tends towards the celerity of light, which is perfectly verified by the observations \cite{Deb95}, \cite{Poi05}.

Modeling is approached using three different visions, Galilean mechanics, the special relativity theory and discrete mechanics. These models lead to the equations corresponding to:
\begin{eqnarray}
\left\{
\begin{array}{llllll}
\displaystyle{ m_0 \: \frac{d  \mathbf V }{dt} =  m_0 \: \mathbf g } \\  \\
\displaystyle{ \frac{d ( m_r \: \mathbf V )}{dt} =  m_0 \: \mathbf g } \\  \\
\displaystyle{ \frac{d  \mathbf V}{dt} = - \nabla \phi +  \mathbf g   } 
\end{array}
\right.
\label{newton}
\end{eqnarray}
where $ m_0 $ is the mass at rest and $m_r$ the mass in motion, $m_r = \gamma \: m_0$ where $\gamma$ is the Lorentz factor $ \gamma = 1 / \sqrt {1 - u^2 / c^2}$; the component of the velocity on the rectilinear axis $x$ is equal to $u = \mathbf V \cdot \mathbf e_x$.

The Newtonian model leads to a velocity that increases linearly as a function of time, and energy tends towards infinity. The mechanical balance described by relativity requires that the moving mass increases indefinitely to conform to Newton's law. The equation of the discrete motion reflects the mechanical equilibrium where the energy associated with the external acceleration $\mathbf g$ is accumulated in the potential $\phi$, and the acceleration becomes null while the velocity becomes constant. While these three visions are strictly disjointed, they are nonetheless complementary in that they propose a version that does not question the previous ones in their field of application.

\textcolor{blue}{\section{Relativistics dynamics  } }

The problem posed corresponds to an accelerated motion where the particle is subjected to a constant acceleration equal to $ \mathbf g $ in a direction defined by a rectilinear axis $Ox$ of unit vector $\mathbf e_x$ that is $g = \mathbf g \cdot \mathbf e_x$; this quantity can also be written $\mathbf g = \nabla \phi_g $ where $\phi_g$ is a scalar potential. This very simple movement makes it possible to highlight the characteristic features of the Newtonian and relativist theories.
\begin{table}[!ht]
\small
\begin{center}
\begin{tabular}{|c|c|c|c|c|c|}   \hline
                   &  velocity $u$      &    acceleration $\gamma$  &   divergence $\delta$&    function  $\beta$   \\ \hline  \hline 
 newtonian             &  $g \: t$  &   $ g$             &  $\displaystyle{\frac{1}{t}}$ &    $  \displaystyle{  \ln  \: t }$             \\ \hline 
 relativity     &  $\displaystyle{\frac{g \: t}{\sqrt{1+\displaystyle{\frac{g^2 \: t^2}{c^2}}}}}$  &   $  \displaystyle{ \displaystyle{\frac{g }{\left( 1+ \displaystyle{\frac{g^2 \: t^2}{c^2} } \right)^{3/2} }} }$    &  $\displaystyle{  \frac{1 }{ \left( 1+ \displaystyle{\frac{g^2 \: t^2}{c^2} }\right) \: t } }$  &  $ \displaystyle{ \displaystyle{  \ln \displaystyle{     \frac{g \: t }{   c \: \sqrt{1+ \displaystyle{\frac{g^2 \: t^2}{c^2} } }  }      } } } $             \\ \hline 
  $\tanh$  law          &  $ c \: \tanh  \left( \displaystyle{ \frac{ g \: t }{ c }} \right)$  &   $g \: \left( 1 - \tanh^2  \left( \displaystyle{ \frac{ g \: t }{ c }} \right) \right)$      &  $\displaystyle{\frac{g \: \left( 1- \tanh^2  (g \: t / c) \right)}{c \: \tanh  (g \: t / c)}}$ &    $  \ln  \left( \tanh  \left( \displaystyle{ \frac{ g \: t }{ c }} \right) \right)$             \\ \hline 
 ${\rm erf}$  law          &  $ c \:{\rm erf} \!\left( \displaystyle{ \frac{\sqrt{\pi} g  t}{ 2 c } } \right)$  &   $\displaystyle{\frac{g}{c}} \: \exp \left( \displaystyle{\frac{- \pi \: g^2 \: t^2 }{ 4 \: c^2 } } \right)$      &  $\frac{\displaystyle{ g \: \exp \left(- \pi \: g^2 \:  t^2 / 4 \: c^2 \right)} }{\displaystyle{c \: {\rm erf} \!\left( \sqrt{\pi} g \: t / 2 c \right) } }$ &    $  \ln \! \left( {\rm erf} \!\left( \displaystyle{ \frac{\sqrt{\pi} g  t}{ 2 c } } \right) \right)$             \\ \hline 
\end{tabular}
\caption{ \it Laws of variation over time of velocity, acceleration, divergence of velocity and $\beta$, the integral of divergence.   }
\label{potar}
\end{center}
\end{table}
\normalsize

Before applying the concepts of discrete mechanics to the problem, it is useful to show the variations in the magnitudes of the problem in these two concepts, velocity $u(t)$, the acceleration acquired by the particle $\bm \gamma(t)$, the divergence of velocity $ \nabla \cdot \mathbf V(t)$ and the function $\beta(t)$, the integral in time of the divergence. Table (\ref{potar}) summarizes the solutions obtained in Newtonian and relativistic mechanics. Two other solutions have been added that are not derived from any transformation and that do not leave the Maxwell equations invariant. These solutions do however satisfy several properties resulting from the Lorentz transformation, velocity tends towards the celerity of light when the time increases indefinitely and, moreover, the behavior of all the quantities corresponds well to Newtonian mechanics when velocity is weak in relation to celerity. The first is a hyperbolic tangent law and the second is an $\rm erf$ error function. The choice of these two functions is arbitrary, but the evolutions of the different associated quantities may lead us to non-obvious conclusions {\it a priori}.

\begin{figure}[!ht]
\begin{center}
\includegraphics[width=7.5cm,height=5.5cm]{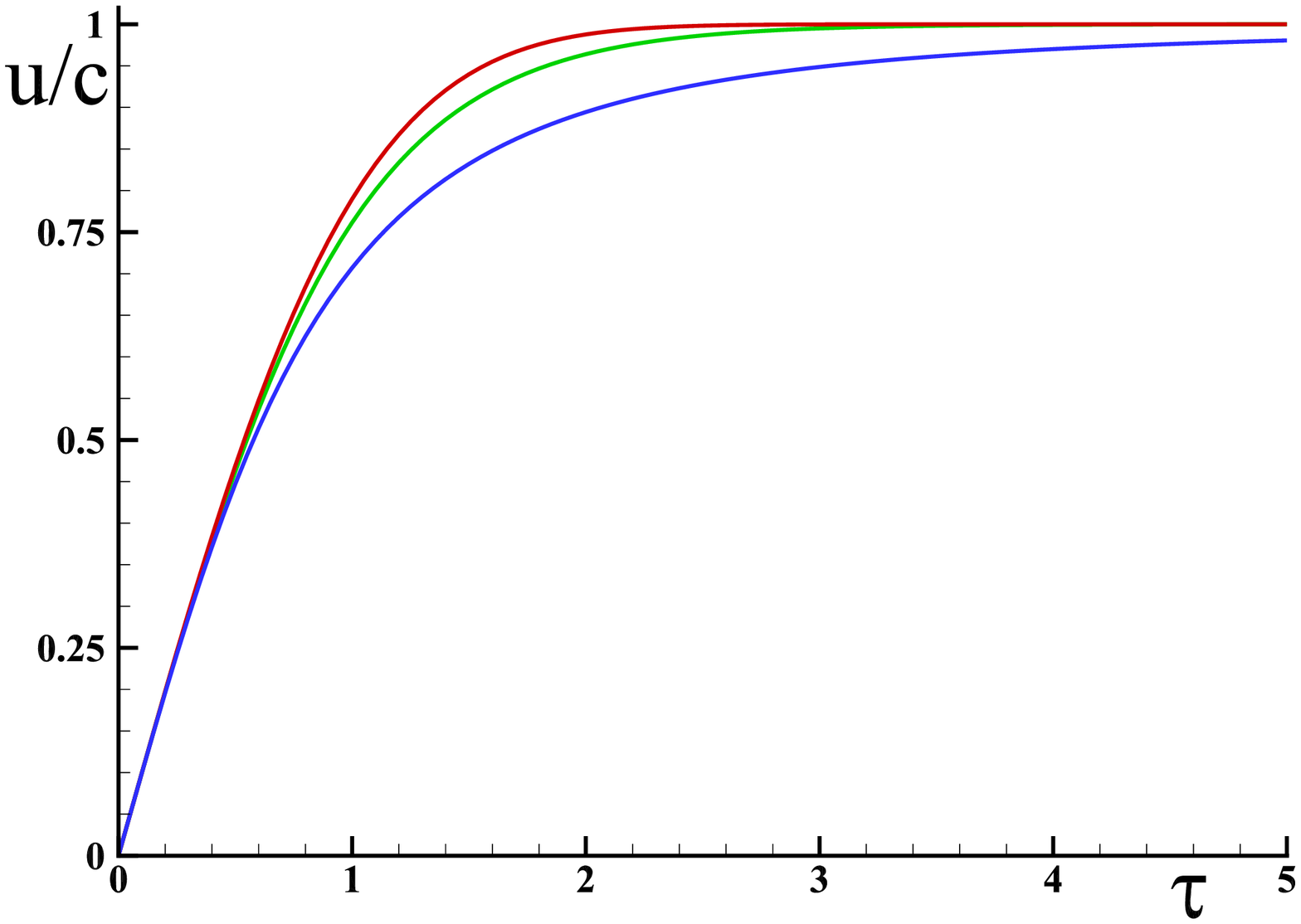}
\includegraphics[width=7.5cm,height=5.5cm]{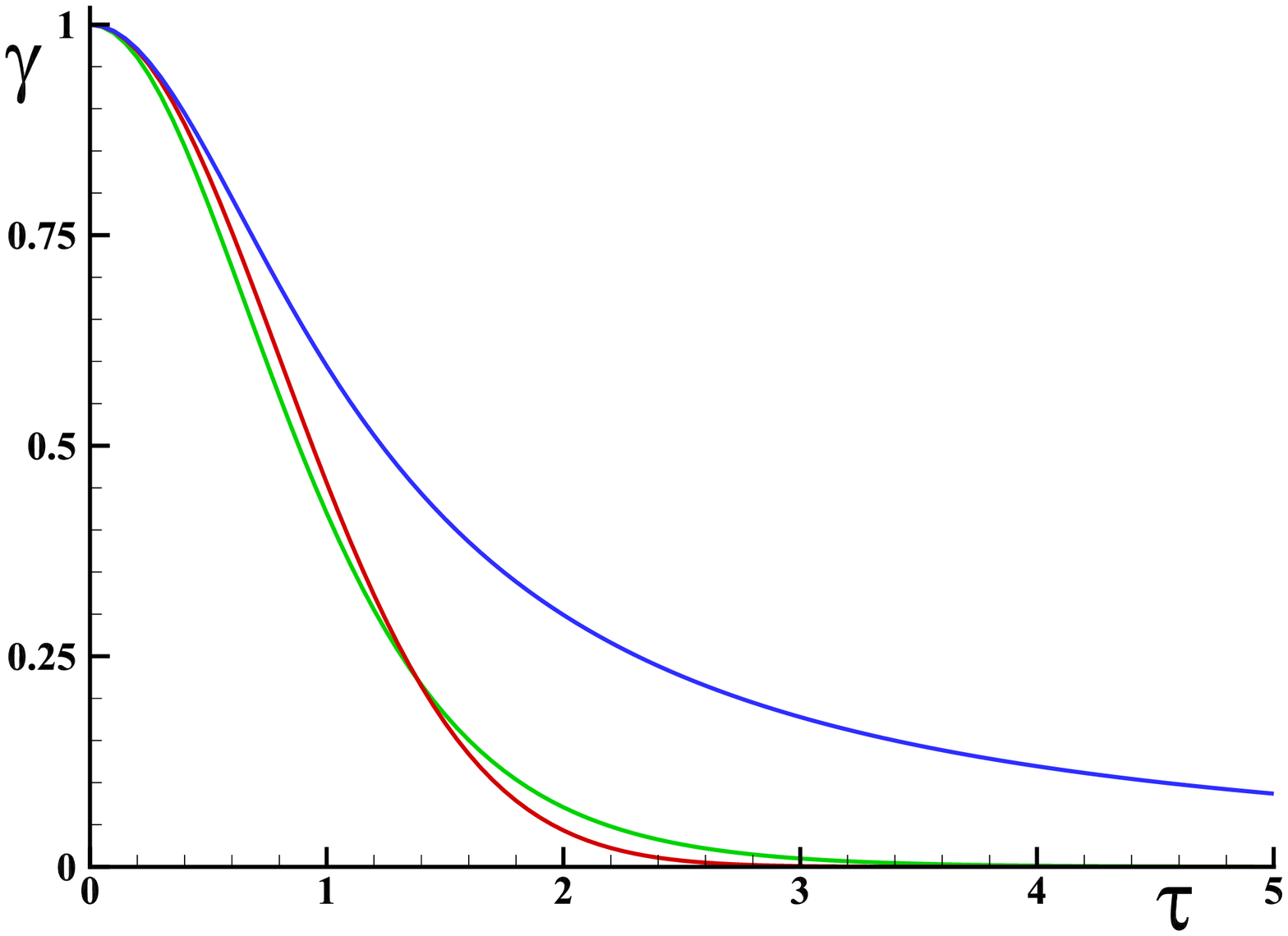}
\includegraphics[width=7.5cm,height=5.5cm]{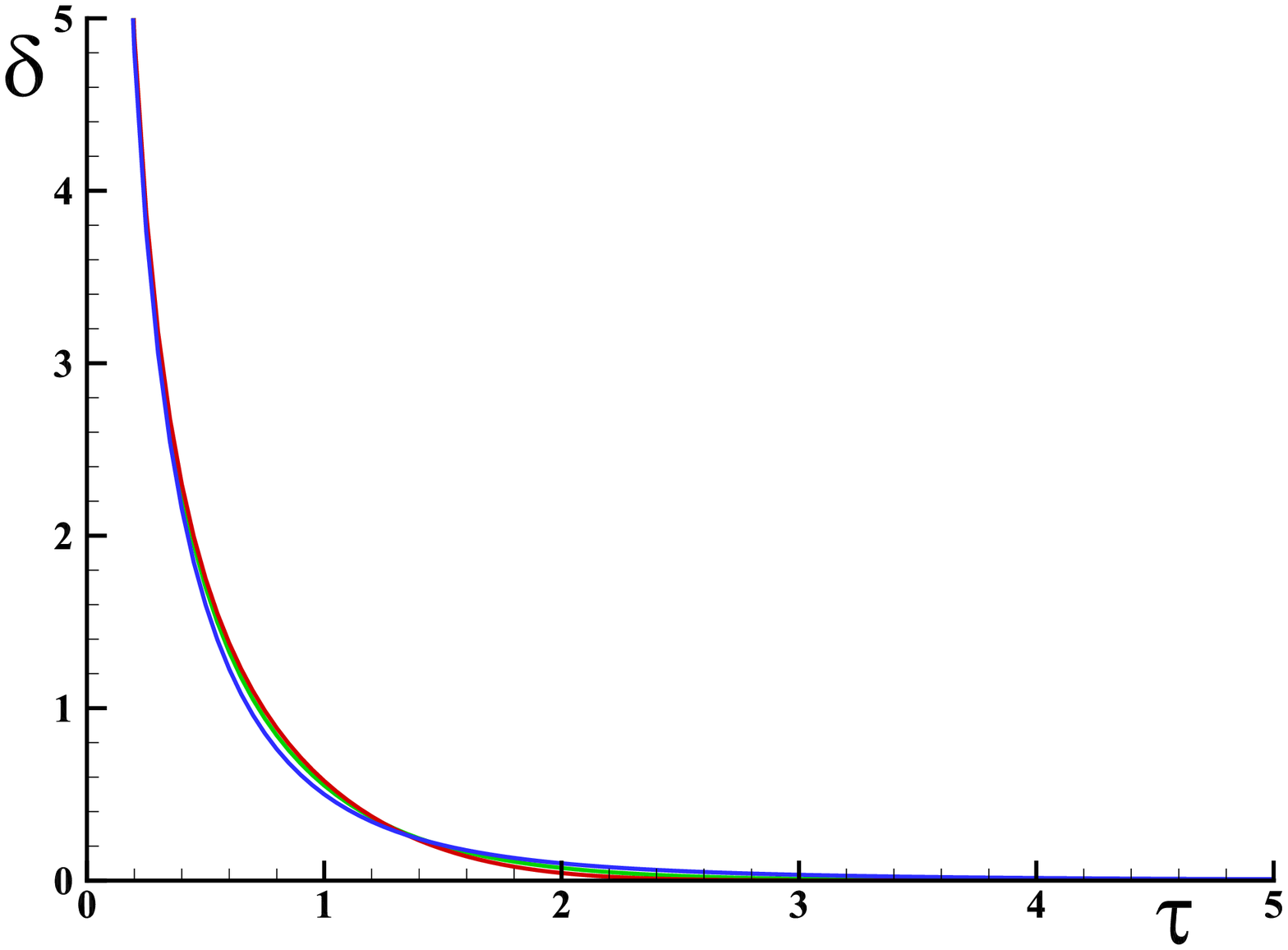}
\includegraphics[width=7.5cm,height=5.5cm]{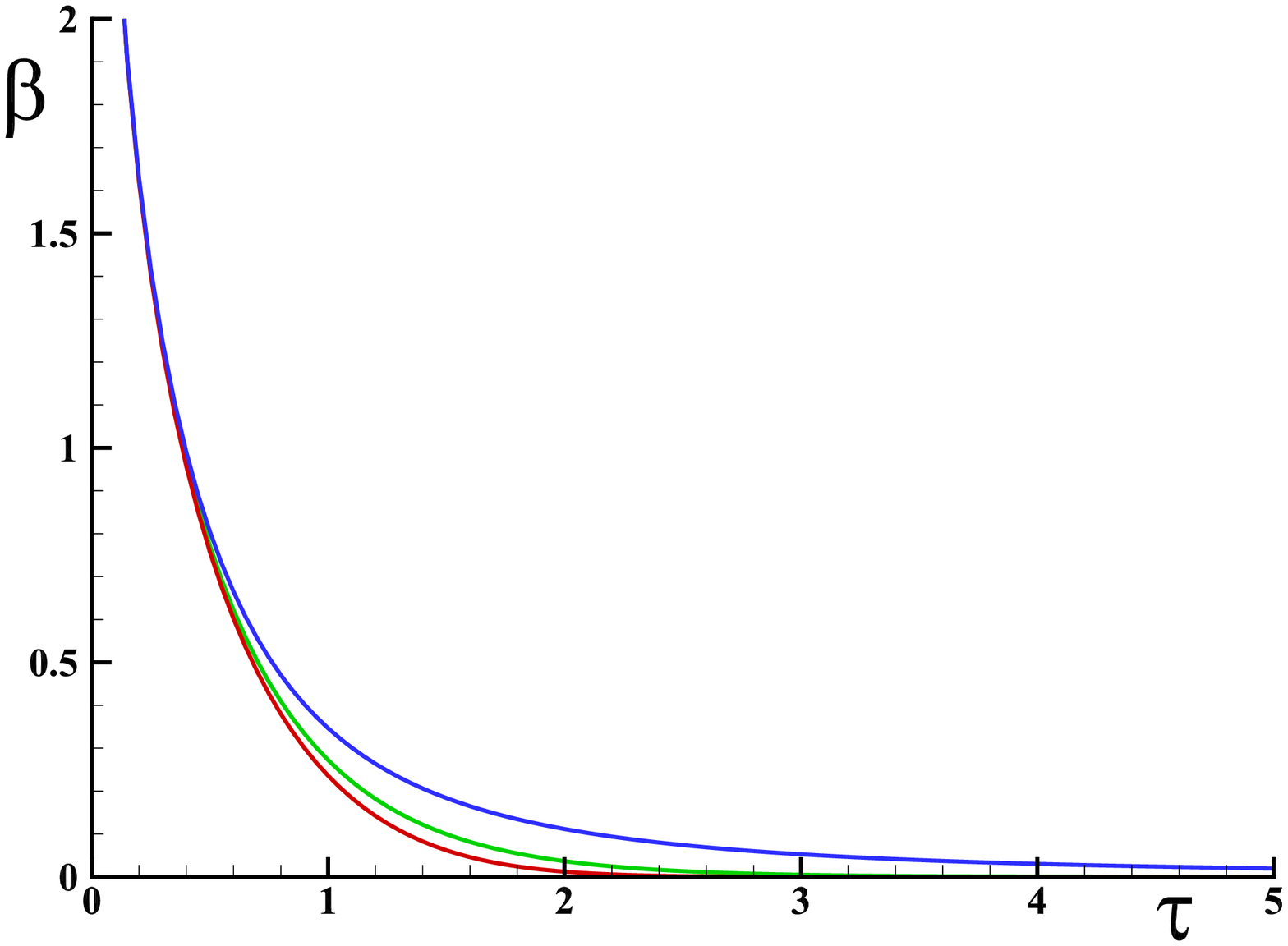}
\caption[step]{\it  Evolutions as a function of reduced time, $\tau = g \: t / c$, of the reduced velocity $u / c$, of the reduced acceleration $\bm \gamma = d (u / c) / dt$, of the divergence of the reduced velocity $\delta = \nabla \cdot (u / c)$ and the function $\beta$. In blue are the evolutions in relativistic mechanics, in green those of the hyperbolic tangent law and in red those of the error function. }
\label{relat}
\end{center}
\end{figure}

The evolutions as a function of reduced time $\tau = g \: t / c$, of the different quantities are illustrated in figure (\ref{relat}). The case of the Newtonian solution is not represented because it is not valid in the context of a relativistic velocity; indeed, the solution in velocity is $u(t) = g \: t$, the particle accumulates all the available energy and the acceleration is always equal to $g$.

The relativistic mechanics of the theory of special relativity is based on the Lorentz transformation; it serves to simply calculate the velocity and the acceleration of a particle subjected to a constant acceleration given in table (\ref{potar}). The two synthetic solutions chosen have a behavior similar to that due to relativity; the velocity increases just a little more quickly to converge to celerity $c$; the acceleration of the particle becomes zero in all cases.
We may be interested in the divergence of velocity when we follow the particle during its movement. Its expression in Lagrangian coordinates is equal to $\nabla \cdot \mathbf V = ( 1/u ) \: du/dt$ to a one-space dimension. The variations in this quantity for the two added laws are very close to the one obtained in relativistic mechanics: it decreases very rapidly towards zero. This quantity, exclusively kinematic linked to the movement, is of primary importance in mechanics, even when one is interested in the motions of particles of different masses and natures.

The last column in table (\ref{potar}) is the time integral of the velocity divergence in Lagrangian coordinates named $\beta$, a dimensionless quantity. Note that the expressions of $\beta$ are logically $\ln (u / c)$ in the last three cases considered. This result is due to the definition of divergence, and is of particular importance that will be highlighted later.

Special relativity is mainly based on the Lorentz transformation which makes Maxwell's equations invariant, the limitation of velocity to that of light in the vacuum, and the principle of relativity which expresses that all the laws of physics are the same in all inertial frames of reference. This set of concordant links led Einstein to present his theory of relativity in 1905, the results of which were confirmed in cosmology or in particle accelerators.

But the fact that the results of incontestable observations are obtained again does not make any theory an irrefutable truth. It is also necessary for the theory to be coherent, to be logical with itself. In particular, it must satisfy energy conservation.

\textcolor{blue}{\section{Discrete physics} }

\textcolor{blue}{\subsection{The physics of the phenomenon} }

The physics of the acceleration of a particle subjected to an external acceleration $\mathbf g$ is not trivial. It is not limited to the application of simple laws, often derived from postulates.

In Newtonian mechanics, the second law $\mathbf F = m \: \bm \gamma$ establishes the link between a force and the acceleration of the body that it provokes by associating it with the mass $m$. In relativistic mechanics this law is supposed to be valid, which induces the necessity of not considering the rest mass $m_0$, but $m_r$, a mass in motion. This somewhat outdated presentation is now replaced by considerations on the momentum $\mathbf q = m \: \mathbf V$, the product of the mass moving by velocity. However, the nature of the problem is not modified. Newton's law remains valid in the theory of relativity.

It is then necessary to separate incontestable facts from repeated experiments with laws that have been widely accepted as representative for centuries. The principles of equivalence and relativity outlined by Galileo have been subjected to direct measurements and cannot be disputed. Likewise, the celerity of light in a vacuum is a constant equal to $c_0$, it is an invariant property of the medium.
It is different from the laws extracted from these principles, which can reproduce the observations but be limited in a well-defined field of application; this is the case, for example, for Newtonian mechanics, representative of movements at low velocities compared to the celerity of light.

Consider, then, a particle with or without mass subjected to a constant external acceleration $\mathbf g$. The velocity of a photon will therefore not always be equal to the celerity of light as in special relativity; When accelerated, the photon will increase in velocity to equal $c$. In the presented case the acceleration $\mathbf g$ applied is directional, $g = \mathbf g \cdot \mathbf e_x$, and the particle will not deviate from its rectilinear trajectory. The only variables of the problem are thus the acceleration of the particle $\bm \gamma$, the velocity $u = \mathbf V \cdot \mathbf e_x$ and its energy per unit of mass denoted $\phi^o$, all depending on space and time. However, velocity and displacement are secondary quantities determined by knowledge of the acceleration and  their values at a fixed instant, for example the initial moment when the velocity and displacement are zero. Acceleration can be expressed from time or distance, since $x = x(t)$.

We are thus led to look for the function $\bm \gamma(t)$; in Newtonian mechanics we have $m_0 \: \bm \gamma = m_0 \: \mathbf g$ and in relativistic mechanics $d (m_r \: \mathbf V) / dt = \mathbf g$. The two laws are debatable, the first because the velocity becomes infinite when time increases and the second because it is the apparent mass which becomes infinite if velocity is limited to celerity; the mass continues to increase while the velocity becomes constant. This is a direct consequence of the supposed validity of Newton's law.

Therefore, it is the second law of Newton itself that is necessary to question. The Fundamental Principle of Dynamics is an instantaneous law that does not translate mechanical equilibrium when force and acceleration become null. It is not a "persistent law" in the sense that the mechanical equilibrium is not ensured intrinsically; it is lacking the notion of energy which would make this relation a law of conservation.

The proposed physical explanation of the phenome-non is as follows: during its motion, the particle accelerates under the effect of $ \mathbf g$, and its energy, set as zero at the initial moment, also increases. The increase in velocity is characterized by its divergence, $\nabla \cdot \mathbf V$; this quantity represents the relative variation in the velocity, which can only decrease if the movement is rectilinear. Two neighboring particles on the same trajectory diverge from each other but their distance cannot increase indefinitely, the movement becomes "incompressible"; this adjective is attached to the movement and not to the constituting matter of the particle, which can be massless. The divergence of the velocity becomes zero, as does the acceleration of the particle. For this persistent mechanical equilibrium, the problem is to determine the velocity of the particle and its energy.

The notion of incompressibility seen as a consequence of the growth of time may be surprising, but it is an everyday reality. A slow movement in the water is considered incompressible even though a shock on its surface produces acoustic waves that propagate in the medium at a velocity equal to the celerity of the water. It is in fact the product $dt \: c^2$ which translates whether the motion is compressible or not, whatever the value of the celerity of the acoustic or light wave; if this product is less than unity, the movement may be considered "compressible". This is the case for the propagation of light if we consider characteristic time periods of the order of $dt \approx 10^{-17} \: s$. When the time increases the movement becomes incompressible and the velocity of the particle is limited to the celerity of the medium; it is an extension of Hugoniot's theorem in fluid mechanics. In particle physics, the acceleration of these particles makes it possible to reach velocities close to the celerity of light.

What remains is to quantify the energy acquired by the particle during the movement; this is defined to within a constant, which is null if one considers its velocity, and zero also at the initial moment. Even if the imposed acceleration $\mathbf g$ is constant, the energy transferred to the particle is limited by the fact that its acceleration $\bm \gamma$ becomes zero. The total energy per unit of mass acquired throughout the motion $\Phi$ must be equal to $c^2$, that is, $e = m \: c^2$, the result predicted by the theory of relativity.

There is therefore a very simple way of deciding on the coherence of the theory of special relativity and the relevance of the Lorentz transformation. It is sufficient to know whether the energy is conserved at the end of the acceleration phase.

\textcolor{blue}{\subsection{Discrete motion equation} }

Discrete mechanics has been the subject of preliminary work \cite{Cal19a} in order to unify the mechanics of solids and incompressible or compressible fluids with or without shock waves; it has also been extended to electromagnetism \cite{Cal18ex}. It reproduces, with the same formalism, the results deduced from the Navier-Lam{\'e}, Navier-Stokes and Maxwell equations and those from other domains such as phase change transfer, porous media, etc.

The law of motion in discrete mechanics is written in the form of a Hodge-Helmholtz decomposition of acceleration:
\begin{eqnarray}
\displaystyle{ \bm \gamma  = - \nabla \phi + \nabla \times \bm \psi + \mathbf g } 
\label{loidis}
\end{eqnarray}

The equation of discrete motion is derived from the conservation equation of acceleration by expressing the deviations of the potentials $\phi$ and $\bm \psi$ as a function of velocity $\mathbf V$. These deviators are obtained on the basis of the physical analysis of the storage-destocking processes of compression and shear energies; the first is written as the divergence of velocity and the second as a dual rotational velocity. The physical modeling of these terms is developed in a work on discrete mechanics \cite {Cal19a}.

The vectorial equation of the movement and its updates are written
\begin{eqnarray}
\left\{
\begin{array}{llllll}
\displaystyle{ \bm \gamma = - \nabla \left( \phi^o - dt \: c_l^2 \: \nabla \cdot \mathbf V \right) + \nabla \times \left( \bm \psi^o - dt \: c_t^2 \: \nabla \times \mathbf V \right)   + \mathbf g } \\  \\
\displaystyle{  \phi^o - dt \: c_l^2 \: \nabla \cdot \mathbf V \longmapsto \phi^o  } \\ \\
\displaystyle{   \bm \psi^o - dt \: c_t^2 \: \nabla \times \mathbf V \longmapsto \bm \psi^o } \\ \\
\displaystyle{\mathbf V^o + \bm \: \gamma \: dt \longmapsto \mathbf V^o  }  \\  \\
\displaystyle{\mathbf x^o + \mathbf V^o \: dt \longmapsto \mathbf x^o  }
\end{array}
\right.
\label{discrete}
\end{eqnarray}

In one dimension of space, the rotational term disappears and the second member remains a term translating the effects of compressibility and the source term $\mathbf g$. In Lagrangian coordinates the term of inertia is absent and $d u / d t \equiv \partial u / \partial t$. By performing the variable change $\partial u / \partial x = \partial / \partial t \: ( dt / dx ) = ( 1 / u ) \: \partial / \partial t$ we get:
\begin{eqnarray}
\left\{
\begin{array}{llllll}
\displaystyle{ \frac{d u}{d t}   = - \frac{1}{u} \: \frac{d }{d t} \left( \phi^o - dt \: c_l^2 \:  \frac{1}{u} \: \frac{d u }{d t}  \right)   +  g } \\  \\
\displaystyle{ \phi^o - \frac{dt \: c_l^2}{u} \: \frac{d u}{d t} \longmapsto \phi^o  }
\end{array}
\right.
\label{discretoneb}
\end{eqnarray}

To our knowledge this equation has no simple analytical solution and must be integrated numerically. Its resolution in Lagrangian coordinates presents multiple problems that are difficult to raise in the presence of an accumulation term on $\phi^o$. In Eulerian coordinates we can use the system (\ref{discrete}) to add the term of inertia $\nabla (\ | \mathbf V \ |^2/2) $, but the variables $x$ and $t$ are not explicitly known; the equation becomes a propagation-diffusion equation. It is then necessary to find the moment when the slope of the function $u (x, t)$ is equal to $g $ to get the $u(t)$ law in Lagrangian coordinates.

\textcolor{blue}{\subsection{Energy conservation } }

Let us consider a particle or a volume element which is followed in its motion induced by external accelerations. In the problem posed, the external accelerations are composed of the force per mass unit imposed, $\mathbf g$ constant and the acceleration associated with the compression (relaxation) of the movement. In an incompressible motion, the divergence of the velocity is zero and $ d \mathbf V / dt$ is always zero on a trajectory, but this is no longer the case for compressible motion; let us take again the definition of the material derivative of the velocity $d \mathbf V / dt$ and that of the temporal derivative:
\begin{eqnarray}
\left\{
\begin{array}{llllll}
\displaystyle{  \frac{d \mathbf V}{d t} =  \frac{\partial \mathbf V}{\partial t}  + \mathbf V \cdot \nabla \mathbf V }  \\  \\
\displaystyle{  \frac{\partial \mathbf V}{\partial t} = - \nabla \cdot \left( \mathbf V \otimes \mathbf V \right) = - \mathbf V \: \nabla \cdot \mathbf V - \mathbf V \cdot \nabla \mathbf V }  \\  \\
\displaystyle{  \frac{d \mathbf V}{dt} = - \mathbf V \: \nabla \cdot \mathbf V }
\end{array}
\right.
\label{particulaire}
\end{eqnarray}

The last expression expresses that, if one follows a particle during its movement, the velocity varies according to the divergence of the motion $\nabla \cdot \mathbf V$ induced implicitly by the external actions. This is true for a fluid element that relaxes or compresses but also for an isolated particle. This expression of the particulate derivative is to be compared with the law of conservation of the density or the mass:
\begin{eqnarray}
\displaystyle{  \frac{d m}{dt} = - m \: \nabla \cdot \mathbf V }
\label{masse}
\end{eqnarray}

The law (\ref{masse}) is the equivalent in discrete mechanics of the equivalence between mass and energy in special relativity.
  Indeed, the divergence of velocity is a variation in energy per unit of mass and $dm / dt = ( 1 / c^2 ) \: d e / dt $.
When the velocity increases during the movement, the divergence of the velocity is negative and the mass increases.

Also, one can easily calculate the evolution of the compression energy (expansion) corresponding to a fluid element (where the energy is equal to $- p \: dv$) or to a particle; this energy per unit mass is written in the general case according to the divergence:
\begin{eqnarray}
\displaystyle{  \frac{d \phi^o}{dt } = - c^2 \:  \nabla \cdot \mathbf V }
\label{demo1}
\end{eqnarray}

By replacing the expression of the divergence by its expression in Lagrangian coordinates $\bm \gamma / \mathbf V$ resulting from (\ref{particulaire}), we can connect the energy to the velocity: 
\begin{eqnarray}
\displaystyle{  \phi^o(t) = - c^2 \:  \int \nabla \cdot \mathbf V \: dt = - c^2 \:  \int  \frac{1}{ \mathbf V } \frac{d \mathbf V}{dt} \: dt  =   - c^2 \:  \int  \frac{d \mathbf V}{ \mathbf V } = - c^2 \: ( \ln \mathbf V + A ) = - c^2 \: \beta }
\label{demo1b}
\end{eqnarray}

When $t \rightarrow \infty$, $\mathbf V \rightarrow c$ and $\ln \mathbf V + A $ must be null and it comes $\beta = \ln (\mathbf V / c)$, a dimensionless quantity.
 This expression of $\phi^o$ corresponds to the energy accumulated by the particle between two equilibrium states, from the moment $t^o $ to the current time $t$.
When the relaxation is complete, the energy per unit mass acquired by the particle is maximal, its value then is equal to $c^2$.
Thus knowledge of the evolution of the velocity of a particle followed in its course makes it possible to calculate, as an accumulation process, the energy $\Phi$ which has been transmitted to it by the external accelerations: 
\begin{eqnarray}
\displaystyle{  \Phi =  c^2 \: \int_0^{\infty} \beta \: d \tau }
\label{demo3}
\end{eqnarray}

The integral of the expression (\ref{demo3}) can be obtained analytically in the first four cases and its value converges, except of course in Newtonian mechanics. 
 The value of the integral to  $\beta$ over reduced time $\tau$ must be equal to the unit so that the energy per unit of total mass equals $\Phi = c^2$.

Table (\ref{potar}) gives the expressions of $\beta$ obtained for the velocity laws treated, in particular in Newtonian mechanics and in relativity. It is sufficient to know the evolution of the velocity $u(t)$ to obtain the energy that the particle accumulated during its $\Phi$ movement; this is given by table (\ref{energiec}). The synthetic laws lead to overvalued energies, $\Phi = 1,234 \: c^2$ for the hyperbolic tangent law and $\Phi = 1,167 \: c^2$ for the error function law.

The energy is not conserved in special relativity where $\Phi = 1.569 \: c^2$. The value of the energy fixed by this theory when the particle is in stationary motion at velocity $u = c$ is indeed $\Phi = c^2$, but the energy yielded by the external medium to set it in motion at velocity $u(t)$ is greater. Only discrete mechanics leads to the expected result. Since the equation of motion is based on the accumulation of energy in the potential $\phi^o$ during the movement, this is not a surprise.
\begin{table}[!ht]
\begin{center}
\begin{tabular}{|c|c|c|c|c|c|}   \hline
  mechanics        &  energy $\Phi$    \\ \hline  \hline 
 newtonian       &  $\infty$            \\ \hline 
 relativistic       &  $1.569 \: c^2$      \\ \hline 
 discrete          &  $ c^2$   \\ \hline   
\end{tabular}
\caption{ \it Energy per unit mass $\Phi$ which has been transmitted to a massive or massless particle by the external accelerations  from the state of rest at a velocity equal to the celerity of the medium.  }
\label{energiec}
\end{center}
\end{table}

\begin{figure}[!ht]
\begin{center}
\includegraphics[width=8.cm]{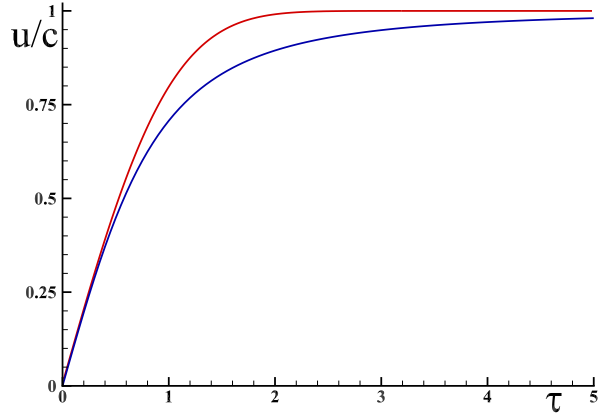}
\caption[step]{\it  Evolutions of the reduced velocity $ u / c $ of a particle with or without mass as a function of reduced time, $\tau = g \: t / c$, in an accelerated rectilinear motion. In blue, the evolution in relativistic mechanics and in red that in discrete mechanics. }
\label{relat-dm}
\end{center}
\end{figure}

Finally, figure (\ref{relat-dm}) shows the result as a function of reduced time obtained on the evolutions of the velocity, respectively in special relativity and in discrete mechanics. The latter is very close to the law based on the error function. The movement defined in relativistic mechanics remains compressible over a longer duration, whereas the law of evolution in discrete mechanics suggests a relatively less compressible movement. As stated in the section on the physics of the phenomenon, the notion of incompressibility depends on the observation time: a movement is less compressible as time increases.

The curve in figure (\ref{relat-dm}) corresponding to relativistic mechanics is a direct emanation of the Lorentz transformation, and the factor $\gamma = 1 / \sqrt {1 - u ^ 2 / c ^ 2} $ is a choice {\it ad hoc}. The overvaluation of energy, of the order of $50$\%, associated with this movement calls into question the validity of this choice. If the final result, where the energy acquired by the particle is $e = m \: \Phi = m \: c^2 $ for the stationary motion is exact, the energy balance associated with the movement as a whole, deduced from velocity $u(t)$, shows an inconsistency. 

\clearpage
\textcolor{blue}{\section{Conclusions} }

The preceding analysis suggests some objective conclusions:

\begin{itemize}
\item the energy of a particle subjected to a constant acceleration is not preserved in the theory of special relativity;
\item The motion equation in discrete mechanics retains energy for any motion of a material medium, but also for isolated particles. It is a law of conservation with continuous memory which ensures the persistence of the essential quantities, the scalar and vector potentials. Mass increases during movement but remains limited.
\end{itemize}

The non-conservation of the energy communicated to a particle by constant acceleration questions the theory of special relativity. Indeed, the total energy yielded by the medium outside the particle of the velocity zero to that of the light is greater than that acquired by the latter, i.e. $e = m \: c^2$. The demonstration of this non-conservation of energy is intrinsic: it relies only on considerations related to the movement itself.

The arguments put forward in relativity, notably the invariance of Maxwell's equations with respect to the Lorentz transformation, are not sufficiently relevant and decisive. While the founding principles of relativity, the principle of equivalence and the relativity of gravitational and inertial effects are indisputable, it would seem that the laws of physics derived from them are not indisputable. Maxwell's equations are incomplete. The equation in discrete mechanics (\ref{discrete}) is from this point of view an alternative that unites electromagnetism \cite{Cal18ex} and mechanics, but also, and coherently, tallies with the results of the theory of relativity. The lasting success of a theory cannot be measured by the yardstick of the restitution of repeated observations, it must also have an internal consistency.



\end{document}